\documentclass[12pt]{article}

\usepackage{cite}
\usepackage{epsf}
\usepackage{graphicx}

\usepackage[english]{babel}
\usepackage{amssymb,indentfirst}
\usepackage{amsmath}

\usepackage{verbatim}

\makeatletter
\renewcommand \thesection {\@arabic\c@section.}
\renewcommand\thesubsection   {\thesection\@arabic\c@subsection.}
\renewcommand\thesubsubsection{\thesubsection\@arabic\c@subsubsection.}
\makeatother

\def\lineup#1{\mbox{$\raise1.0ex\hbox{--} \kern-0.6em#1$}}

\def\starup#1{\mbox{$\raise1.6ex\hbox{$*$} \kern-0.5em#1$}}
\def\starupp#1{\mbox{$\raise1.8ex\hbox{$*$} \kern-1.0em#1$}}
\def\staruppp#1{\mbox{$\raise1.0ex\hbox{$*$} \kern-0.5em#1$}}
\def\sstarup#1{\mbox{\scriptsize $\raise1.8ex\hbox{$*$} \kern-.7em#1$}}

\def\ttildeup#1{\mbox{$\raise0.0ex\hbox{\Large $\; \tilde{}$} \kern-0.45em#1$}}
\def\bbar#1{\mbox{$\raise-0.4ex\hbox{\Large $\; \bar{}$} \kern-0.35em#1$}}

\begin{document}

\title{ 
On a possible manifestation \\ of the four color symmetry $Z'$ boson \\ in $\mu^+\mu^-$ events at the LHC 
%MQLS $Z'-$boson Discovery Reach at LHC                                               
}
\author{A. D. Smirnov\thanks{E-mail: asmirnov@univ.uniyar.ac.ru}, 
Yu. S. Zaitsev\thanks{E-mail: zajax.jr@mail.ru}\\
{\small\it Division of Theoretical Physics, Department of Physics,}\\
{\small\it Yaroslavl State University, Sovietskaya 14,}\\
{\small\it 150000 Yaroslavl, Russia.}}
\date{}
\maketitle
\begin{abstract}
\noindent

The cross section of the $\mu^+\mu^-$ pair production in $pp$-collisions at the LHC 
is calculated with acount of the $Z'$ boson induced by the minimal four color 
quark-lepton symmetry($MQLS$).   
The $\mu^+\mu^-$ invariant mass spectrum with account of the $MQLS$ $Z'$ boson  
is analysed in dependence on the $Z'$ mass. The mass region for the $MQLS$ $Z'$ boson  
observable at the LHC is found in dependence on the significance and on the integrated luminosity. 

\vspace{5mm} \noindent \textit{Keywords:} Beyond the SM; four color
symmetry; Pati--Salam; $Z'$~boson; LHC.

\noindent
PACS number: 12.60.-i

\end{abstract}

The search for a new physics beyond the Standard Model (SM) is an actual field
of the elementary particle physics now. The near start of LHC will essentially 
enlarge the possibilities for search for new physics effects and  
in this situation the extensions of SM predicting new efects at the LHC energies 
(such as supersymmetry, left--right symmetry, two Higgs model, etc.) acquire 
the special interest.     

One of the possible variants of such new physics can be induced by the
possible four color quark--lepton symmetry of Pati--Salam type treating leptons 
as the fourth color \cite{PS}.
The immediate consequence of this symmetry is the prediction 
of the specific new gauge particles -- vector or chiral gauge leptoquarks and 
one or two additional neutral $Z'$~bosons. 
In particular case of the minimal unification with the SM by the gauge group  
\begin{eqnarray}
G_{new} = SU_V(4)\times SU_L(2)\times U_R(1) 
\label{eq:G421}
\end{eqnarray}
\noindent 
(MQLS~model \cite{x0,x1} ) the four color symmetry 
(described here by the vector color group $SU_V(4)$) 
predicts one vector leptoquark $V^\pm$ with electric charge $\pm 2/3$
and one additional neutral $Z'$~boson. 
The interactions of these gauge fields with fermions are defined by 
the group~(\ref{eq:G421}) and give the possibility to investigate quantitatively 
the effects of these paricles in dependence on their masses.  
Below we concentrate our attention on the additional $Z'$~boson predicted by the four color symmetry 
unified with SM by the minimal group~(\ref{eq:G421}).    

It should be noted that additional $Z'$~bosons appear in many extentions of the SM.  
In dependence on the model the most stringent lower limits on $Z'$~masses are of order 
$M_{Z_{\psi}}, M_{Z_{\chi}} > 0.822$ TeV~\cite{PDG08} for $Z'$~bosons of $E_6$~model 
from $Z'$~direct searches at the Tevatron~\cite{AaltonenPRL99}
% LEP \cite{LEPEWWG} and  
and $M_{Z_{LR}}>0.86$ TeV, $M_{Z'_{SM}}>1.5$ TeV~\cite{PDG08}             
for $Z'$~bosons of $LR$~model and of the model with SM coupling constants 
from the global electroweak analysis~\cite{CheungPLB517}. 
The most stringent mass limit for the MQLSM $Z'$~boson  can be extracted 
from the LEP leptonic cross section        
%$\sigma_{\bar f f} = \sigma(e^+ e^- \to \gamma, Z, Z' \to \bar f f)$ 
$\sigma_{\bar l l} = \sigma(e^+ e^- \to \gamma, Z, Z' \to \bar l l)$ 
\cite{x1}
and with account the LEP2 data~\cite{LEPCOLL2006}
is of about 
\begin{equation} 
m_{Z'} > 1.4 \, \, TeV .   %\mbox{ТэВ} .
\label{eq:mZ1s} 
%\nonumber
\end{equation}

It is interesting now to know what is the mass region in which the MQLSM $Z'$~boson can be observable  
at future search for $Z'$ at the LHC. The most clean sign of the $Z'$ boson 
production at the LHC can be the possible excess of $\mu^+\mu^-$ events resulting from 
$Z'$ decays into $\mu^+\mu^-$ pairs. 
The investigation of the $\mu^+\mu^-$ invariant mass spectrum and of its possible 
deviation from the SM prediction will give the possibility to detect 
the production of $Z'$ or to set the new mass limit on its mass. 
So the analysis of the process $pp\rightarrow \gamma,Z,Z' \rightarrow \mu^+\mu^-$ 
at the LHC energy is of interest.

In the present paper we calculate the cross section of lepton-antilepton pair
production in $p p$-collisions at the LHC with account of $Z'$-boson predicted 
by the the four color symmetry of type~(\ref{eq:G421}). 
We obtain and discuss the mass region in which this $Z'$ can be discovered at the LHC 
in comparison with the other models ($E_6,LRM,SSM$ \cite{Leike,rizzo_06}) 
also predicting $Z'$~bosons.         
                
Interactions of the neutral gauge bosons with the SM fermions can be written as 
\begin{equation} 
\label{lagranjian1}
%    \mathcal{L}_{NC}^{gauge}= - |e| \sum_{i=\gamma,Z_1,Z_2}J^i_{\mu}
    \mathcal{L}_{NC}^{gauge}= - |\, e\, | \, \sum_{i}J^i_{\mu}
    A_i^{\mu},
\end{equation}
where $e$ is an electron charge, $A_i \, (i=0, 1, 2)$ denote the gauge
boson $\gamma, Z, Z'$,   
\begin{equation}
 J_\mu^i=\sum_{f_a} \bar{f_a}\gamma_\mu(v_{f_a}^i+\gamma_5 a_{f_a}^i)f_a
\end{equation}
are the neutral currents and $v_{f_a}^i \equiv v_{f_a}^{A_i} $ and $a_{f_a}^i \equiv a_{f_a}^{A_i} $ 
are the vector and axial coupling constants 
of fermion $f_a$  with gauge boson $A_i$ ($f$ is quark or lepton, $a=1, 2$ for up and down fermion).

 In general case the mass eigenstates $Z$ and $Z'$ are superposition 
%\begin{equation} 
%\label{ZZmixing}
%    Z^\mu=Z_1^\mu \cos \theta_m -Z_2^\mu \sin \theta_m, \qquad
%    Z'^\mu=Z_1^\mu \sin \theta_m +Z_2^\mu \cos \theta_m,  
%\end{equation}
of two basic fields $Z_1$ and $Z_2$.  
With account of $Z-Z'$ mixing the coupling constants of $Z$ and $Z'$ with   
fermions can be written as  
\begin{eqnarray} 
\label{vaZ}
&&  v_{f_a}^{Z}=v_{f_a}^{Z_1} \cos \theta_m - v_{f_a}^{Z_2} \sin \theta_m, \qquad
  a_{f_a}^{Z}=a_{f_a}^{Z_1} \cos \theta_m - a_{f_a}^{Z_2} \sin \theta_m, \qquad
\\
\label{vaZprime}
&& v_{f_a}^{Z'}=v_{f_a}^{Z_1} \sin \theta_m + v_{f_a}^{Z_2} \cos \theta_m, \qquad
  a_{f_a}^{Z'}=a_{f_a}^{Z_1} \sin \theta_m + a_{f_a}^{Z_2} \cos \theta_m, \qquad
\end{eqnarray}
where 
\begin{eqnarray}
 v_{f_a}^{Z_1}= \frac{(\tau_3)_{aa}- 4Q_{f_a}s^2_W}{4s_Wc_W}, \,\,\,\,\,\
 a_{f_a}^{Z_1}= \frac{(\tau_3)_{aa}}{4s_Wc_W}
\,\label{eq:thr}
\end{eqnarray}
are the SM coupling constants, $\theta_m$ is  the $Z-Z'$ mixing angle,   
$s_W=\sin \theta_W$, $c_W= \cos \theta_W $, $\theta_W$ is the Weinberg weak mixing angle, 
$\tau_3$ is the Pauli matrix, $Q_{f_a}$ is electric charge of fermion $f_a$ in part of $|\, e\, |$. 

The interaction of $Z_2$ field with fermions depends on the origin of this field.       
In MQLS model the coupling constants 
$v_{f_a}^{Z_2}, \,  a_{f_a}^{Z_2} $ have the form~\cite{PovSmPhAN65}  
\begin{eqnarray}
%v'_{f_a}\approx 
&& v_{f_a}^{Z_2}= \frac{1}{c_W s_S\sqrt{1-s_W^2-s_S^2}}[
c_W^2\sqrt{\frac{2}{3}}(t_{15})_f- (Q_{f_a}-\frac{(\tau_3)_{aa}}{4})s_S^2],
\,\label{three}
\\
%\end{eqnarray}
%\begin{eqnarray}
%a'_{f_a}\approx 
&& a_{f_a}^{Z_2}=
\frac{s_S}{c_W \sqrt{1-s_W^2-s_S^2}}\frac{(\tau_3)_{aa}}{4} 
\label{fourth}
\end{eqnarray}
where $t_{15}$ is the 15-th generator of $SU_V(4)$ group,  
$s_S=\sin \theta_S$ is the parameter of the model defined be relation 
of elecromagnetic and strong coupling constants, $\theta_S$ is a strong mixing angle. 
In MQLS model the $Z-Z'$ mixing angle is small ($\theta_m < 0.006$) and 
we neglect below the $Z-Z'$ mixing believing $Z \approx Z_1$ and $Z' \approx Z_2$.

The differential cross-section of lepton-antilepton pair production in
fermion-antifermion collision in the tree approximation is well
known and for $m_f \ll \sqrt{s}$ can be written as \cite{Leike,rizzo_06} 
\begin{equation}\label{Diff1}
    d \sigma
    (\bar{f_a}f_a \stackrel{\gamma,Z,Z'}{\longrightarrow} l^+ l^-)
    =
    \frac{\pi \alpha^2 s}{2 N_c}
    \sum_{i,j}\left( 2 z B_{f_al}^{ij}+(z^2+1)C_{f_al}^{ij} \right) 
%    \sum_{i,j=\gamma,Z,Z'}\left( 2 z B_{f_al}^{ij}+(z^2+1)C_{f_al}^{ij} \right) 
%    Re(P_i(s) P_j^*(s)) d z.
    P_{ij}(s) d z.
\end{equation}
Here $\alpha$ is the fine structure constant, $s$ is the   
invariant mass square of the initial $\bar{f}f$ pair, $N_c=1(3)$ denote colour factor for leptons
(quarks), $z=\cos{\theta}$, $\theta$ is the scattering angle in the center of mass frame,  
 $B^{ij}_{f_al}, C^{ij}_{f_al}$ are the combinations of the coupling constants of the form 
\begin{eqnarray*}
    C^{ij}_{f_al}=(a_{f_a}^i a_{f_a}^j+v_{f_a}^i v_{f_a}^j)(a_l^i a_l^j+v_l^i v_l^j), \\
    B^{ij}_{f_al}=(a_{f_a}^i v_{f_a}^j+v_{f_a}^i a_{f_a}^j)(a_l^i v_l^j+v_l^i a_l^j), 
\end{eqnarray*}
$P_{ij}(s) = Re(P_i(s) P_j^*(s))$
and $P_i(s) = 1/(s-M_i^2+i M_i \Gamma_i)$ is a factor coming from
propagator of boson $A_i$ with mass $M_i \equiv M_{A_i} $ and with width $\Gamma_i \equiv \Gamma_{A_i}$.

Integration of (\ref{Diff1}) over $z$ gives the total cross-section in the form 
\begin{equation} 
\label{full_ff_cs}
\sigma(\bar{f_a}f_a \stackrel{\gamma,Z,Z'}{\longrightarrow} l^+ l^-)=
\frac{4 \pi s \alpha^2}{3 N_c} 
\sum_{i,j}C^{ij}_{f_al} 
%\sum_{i,j=\gamma,Z,Z'}C^{ij}_{f_al} 
%Re(P_i(s) P_j^*(s)).
P_{ij}(s).
\end{equation}

The differential cross-section of $\mu^+\mu^-$ pair production in
proton--proton collision (Drell--Yan process) with account of an aditional $Z'$-boson 
can be written as 
\begin{align} 
\label{pp_full1}
d\sigma(pp \stackrel{\gamma,Z,Z'}{\longrightarrow} \mu^+ \mu^-)=
\sum_k F_k(x_1,x_2,s) \sigma(\bar{q_k} q_k
\stackrel{\gamma,Z,Z'}{\longrightarrow} \mu^+ \mu^-)dx_1 dx_2 
\end{align}
where  $s=x_1 x_2 S$, $S$ is total energy square of colliding protons, 
$x_{1,2}$ is parton fraction of proton momentum and the function
$F_k(x_1,x_2,s)$ can be expressed in the terms of parton distribution functions 
$f_{q_k}(x,s)$ ($f_{\bar{q}_k}(x,s)$) of $k$-flavor quarks $q_k$ (antiquarks $\bar{q}_k$) as 
\begin{align} 
\label{qq_probab} 
\notag
F_k(x_1,x_2,s)=f_{q_k}(x_1,s)f_{\bar{q}_k}(x_2,s)+f_{\bar{q}_k}(x_1,s)f_{q_k}(x_2,s).
\end{align}

For the further analysis it is useful to go over from the variables $\{x_1,x_2\}$ to
\begin{equation} 
\label{peremennie}
    M^2=x_1 x_2 S, \qquad y=\ln{\frac{x_1}{x_2}} 
\end{equation}
where $M$ is invariant mass of quark-antiquark pair (which is
equal to invariant mass of the final $\mu^+ \mu^-$ pair), $y$ is the final lepton
rapidity, $\sqrt{S}$ is the energy of colliding protons. 
In terms of the variables $\{M,y\}$ the  cross section (\ref{pp_full1}) takes the form 
\begin{align} 
\label{pp_vs_My}
d\sigma(pp \stackrel{\gamma,Z,Z'}{\longrightarrow} \mu^+ \mu^-)=
\frac{8 \pi M^3 \alpha^2}{9 S}\sum_k
F_k(\frac{M}{\sqrt{S}}e^y,\frac{M}{\sqrt{S}}e^{-y},M^2) 
%\times \\
%\notag \times 
%\sum_{i,j=\gamma,Z,Z'}C^{ij}_{q_k\mu}
\sum_{i,j} C^{ij}_{q_k\mu}
% Re\left(\frac{1}{M^2-M_i^2-i M_i \Gamma_i}\frac{1}{M^2-M_j^2+i M_j \Gamma_j}
% Re\left( P_i(M^2)  P^*_j(M^2) \right) 
P_{ij}(M^2)
dM dy.
\end{align}

Integration of the cross section (\ref{pp_vs_My}) over $y$ gives
the $\mu^+\mu^-$ invariant mass spectrum in the form 
\begin{align} 
\label{pp_vs_M}
\frac{d\sigma(pp \stackrel{\gamma,Z,Z'}{\longrightarrow} \mu^+
\mu^-)}{dM}=\frac{8 \pi M^3 \alpha^2}{9 S} 
%\sum_k I_k(M,S) 
%%\times \\
%%\notag \times 
%%\sum_{i,j=\gamma,Z,Z'}C^{ij}_{q_k \mu}
%\sum_{i,j}C^{ij}_{q_k \mu}
\sum_{i,j} I^{ij}(M,S) 
%% Re\left( \frac{1}{M^2-M_i^2-i M_i \Gamma_i}\frac{1}{M^2-M_j^2+i M_j \Gamma_j} \right)
P_{ij}(M^2)
\end{align}
where
\begin{equation} 
\label{integralij}
I^{ij}(M,S) = \sum_k I_k(M,S) C^{ij}_{q_k \mu}, 
\end{equation}
\begin{equation} 
\label{integral}
I_k(M,S)=\int_{-\ln{\sqrt{S}/M}}^{+\ln{\sqrt{S}/M}}
F_k(\frac{M}{\sqrt{S}}e^y,\frac{M}{\sqrt{S}}e^{-y},M^2) \, dy.
\end{equation}

The analysis of the possible new effects in $\mu^+\mu^-$ invariant mass spectrum at the LHC 
was performed by CMS collaboration \cite{CMS_1,CMS_2}. This analysis is based on 
the statistical significance $\mathcal{S}'$ estimation of the signal
$pp\rightarrow\gamma,Z,Z'\rightarrow\mu^+\mu^-$ events in the presence of the 
background $pp\rightarrow\gamma,Z\rightarrow\mu^+\mu^-$ events, with $Z'$
predicted by some models ($E_6, LRM, ALRM, SSM$).

There are different definitions of significance estimators in literature 
%(see \cite{Exp_sig_obs} for example) 
and we use here the next one~\cite{Exp_sig_obs}  
\begin{equation} 
\label{Significance}
%    \mathcal{S}'=\sqrt{2\left((N_s+N_b)\ln{\left(1+\frac{N_s}{N_b}\right)}-N_s\right)}
    \mathcal{S}'=\sqrt{2\left[ (N_s+N_b)\ln{\left( 1+N_s/N_b \right)} - N_s \right]}
\end{equation}
where $N_s$ and $N_b$ are number of signal and background events in
the dilepton invariant mass region $M_{Z'}\pm \Delta M$. 
The numbers $N_s$ and $N_b$ can be calculated from  the cross sections of type (\ref{pp_vs_M}) as 
\begin{equation} 
\label{sigmas}
N_s=L \,\sigma_s(M_{Z'},\Delta M), \qquad  
\sigma_s(M_{Z'},\Delta M)=\int_{M_{Z'}-\Delta M}^{M_{Z'}+
\Delta M}\frac{d\sigma(pp \stackrel{\gamma,Z,Z'}{\longrightarrow} \mu^+ \mu^-)}{dM}dM,  
\end{equation}
\begin{equation} 
\label{sigmaSM}
N_b=L \, \sigma_{SM}(M_{Z'},\Delta M),  \qquad 
\sigma_{SM}(M_{Z'},\Delta M)=\int_{M_{Z'}-\Delta M}^{M_{Z'}+
\Delta M}\frac{d\sigma(pp \stackrel{\gamma,Z}{\longrightarrow} \mu^+ \mu^-)}{dM}dM  
\end{equation}
where $L = \int \mathcal{L} dt$ is the integrated luminosity,    %  $\sigma_{s(SM)}(M_{Z'},\Delta M)$ 
$\sigma_{s}(M_{Z'},\Delta M)$ and $\sigma_{SM}(M_{Z'},\Delta M)$ are  
the cross sections for the signal and background events in the mass region $M_{Z'}\pm\Delta M$. 
Here $\Delta M$ is a mass window chosen below as $\Delta M=0.85 \, \Gamma_{Z'}$, which 
corresponds to the width of $2 \sigma$ in the case of Gaussian distribution \cite{Landsberg}.

The $\mu^+\mu^-$ invariant mass spectrum (\ref{pp_vs_M}) depends on the mass $M_{Z'}$  
and on the width $\Gamma_{Z'}$  of $Z'$ boson. The fermionic decays of $Z'$ boson 
are defined by the coupling constants~(\ref{vaZprime}--\ref{fourth}) and   
the corresponding partial widths of $Z'$ boson decays to $f_a \bar{f_a}$ pairs 
for $m_{f_a} \ll M_{Z'}$ have the form 
\begin{equation}
\label{GamToFermion}
\Gamma(Z'\rightarrow f_a \bar{f_a})=N_f M_{Z'}\frac{\alpha}{3} \big((v_{f_a}^{Z'})^2+(a_{f_a}^{Z'})^2 \big)
\end{equation}
where the color factor $N_f=1(3)$ for leptons(quarks) $f=l(q)$. 

In the case of the Higgs mechanism of the fermion mass generation the four color symmetry 
of type~(\ref{eq:G421}) in addition to the SM Higgs doublet predicts the new scalar doublets: 
the color octet of scalar gluon doublets $F_{j a},j=1,\ldots,8$, two color triplets
of scalar leptoquark doublets $S^{(\pm)}_{a\alpha}, \alpha=1,2,3$ 
and additional colorless scalar doublet $\Phi'_a$ as well as some other scalar fields 
(more details can be found in~\cite{x2}). 
The analysis of the mass limits for the scalar doublets $F_{a}$, $S^{(\pm)}_{a}$, $\Phi'_a$   
from $S, \, T, \, U$ parameters of electroweek radiative corrections~\cite{ADPLB531,PovSmPhAN66},  
from $K^0_L \to e^{\mp} \mu^{\pm}$ and $B^0 \to e^{\mp} \tau^{\pm}$ decays~\cite{ADMPLA22,ADPhAN71,PDG08},  
from the magnetic moments of muon~\cite{PovPhAN69} and of neutrino~\cite{PovPhAN70} 
%from the decays of type $b \to s \gamma, \, \mu \to e \gamma$~\cite{xxxx}  
showed that these scalar doublets can be light, with mases below 1~TeV. 
So, the MQLSM $Z'$ boson can decay also into pairs of these scalar particles, 
which gives the additional to~(\ref{GamToFermion}) contributions to $Z'$ width.              

Writing the interaction of $Z'$ boson with scalar field $\Phi$ as    
\begin{equation}
\mathcal{L}_{Z' \Phi \Phi}= i g_{Z' \Phi \Phi} Z'_{\mu}
\left(\partial^{\mu}\Phi^{*}\Phi - \Phi^{*} \partial^{\mu}\Phi
\right) 
\end{equation}
where $g_{Z' \Phi \Phi}$ is the corresponding coupling constant  
for the width of $Z'$ boson decay into $\Phi \tilde{\Phi}$ pair we have the expression  
\begin{equation} 
\label{ScalarPartialWidth}
\Gamma (Z' \rightarrow \Phi \tilde{\Phi})= N_{\Phi} M_{Z'} \frac{  g_{Z'\Phi \Phi}^2 }{48\pi} 
\left(1-\frac{4 m_{\Phi}^2}{M^2_{Z'}}
\right)^{3/2} 
\end{equation}
where $N_{\Phi}$ is the color factor ($N_{F_a}=8$ for scalar gluons, $N_{S^{(\pm)}_a}=3$ for
scalar leptoquarks, $N_{\Phi_a'}=1$ for the additional colorless scalar doublet) 
and $m_{\Phi}$ is a mass of the scalar particle. 

The scalar gluons $F_{a}$ and the scalar leptoquarks $S^{(\pm)}_{a}$ gives the main 
contribution into $Z'$ boson width of type~(\ref{ScalarPartialWidth}). 
The coupling constants of these particles with $Z'$ boson are predicted by the MQLS model as 
\begin{align} 
\label{couplings}
g_{Z'F_aF_a}=-\frac{e}{2}\frac{\sigma}{s_W c_W},\quad
g_{Z'S^{(\pm)}_a S^{(\pm)}_a}=-e\left(\frac{\sigma}{2s_Wc_W}\pm\frac{2 t_W}{3\sigma}\right)
\end{align}
where  $t_W = \tan \theta_W$ and  $\sigma=s_W s_S/\sqrt{1-s_W^2-s_S^2}$. 

The partial widths of $Z'$ decays into scalar gluon and scalar leptoquark pairs 
of type~(\ref{ScalarPartialWidth}), (\ref{couplings}) depend on parameter $s_S$ and on masses 
of these scalar particles and on $Z'$ mass $M_{Z'}$. 
The parameter $s_S$ is defined by the mass scale $M_c \sim M_V$ of the four-color symmetry breaking 
and by the intermediate mass scale $M'\sim M_{Z'}$ \cite{x1,x2}. 
For example for $M' \sim  10 \, TeV$ and for $M_c = 10^4 \, TeV, \, 10^6 \, TeV, \, 10^8 \, TeV$ 
we have $s_s^2 = 0.070, \, 0.112, \, 0.154$ respectively~\cite{x2}.   
For numerical estimations we use below the value $s_s^2=0.114$ which corresponds to 
$M_{Z'} \sim 1 - 5 \, TeV$ and $M_c \sim 10^3 \, TeV$.      
The current mass limits for scalar leptoquarks are of about  $m_{S^{(\pm)}}\gtrsim 250-300 \, GeV$ 
in dependence on details of their interactions with fermions~\cite{PDG08}.   
At the present time there are no reliable experimental mass limits for the scalar gluons but it is reasonable 
to expect their masses to be of the same order or slightly greater than the scalar leptoquark ones.  
Below in calculations we use for the masses of the scalar leptoquarks and of the scalar gluons     
the value $300$ GeV.   
With these values of $s_s^2$ and of the masses of scalar particles 
the relative total width of $Z'-$boson $\Gamma_{Z'}/M_{Z'}$ occurs to be equal  to 
\begin{equation} 
\label{gammaZ1}
\Gamma_{Z'}/M_{Z'} = 4.3 \, \% \; (1.1 \, \% , \;  3.2 \, \% ), \;\; 5.2 \,\% \; (2.0 \,\% , \;  3.2 \,\% ),
\;\;  5.3 \,\% \; (2.1 \,\% , \;  3.2 \,\%) 
\end{equation}
for $M_{Z'}$ of about respectively 1 TeV, \, 3 TeV, \, 5 TeV and above, 
the corresponding values of the relative widths
of the $Z'$~decays respectively into scalar particles and into fermions are shown in papenthesis.

\begin{figure}[htb]
\vspace*{0.5cm}
\centerline{\epsfxsize=0.8\textwidth\epsffile{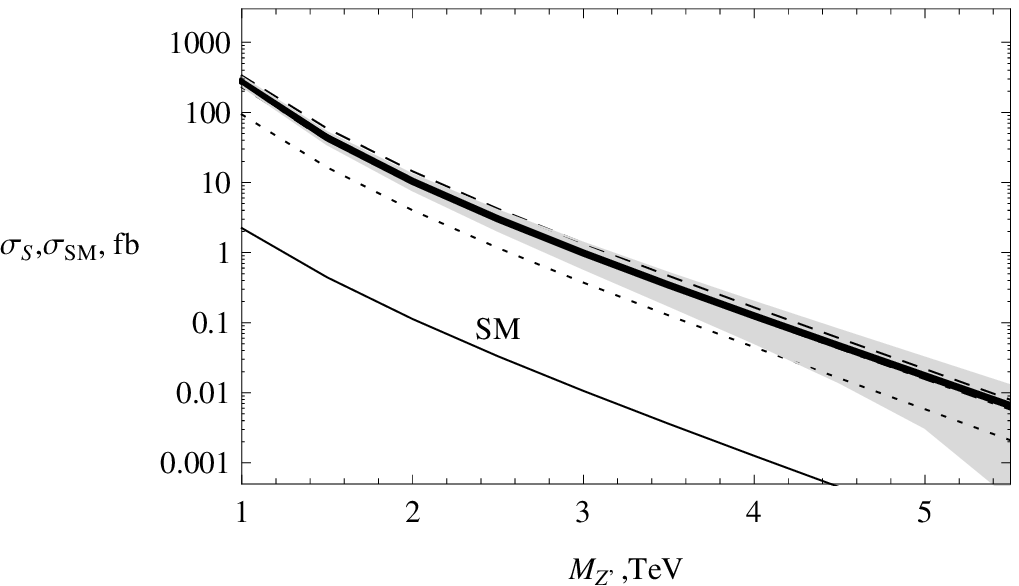}}
%\vspace*{1mm}
\caption{
%Cross section $\sigma_{s}(M_{Z'},\Delta M)$ of $\mu^+\mu^-$ pair production at the LHC  
%as a function of $Z'$ boson mass $M_{Z'}$  
%in  $MQLS$ model (solid line), in  $E_6(\psi)$ model (dotted line), 
%in $LRM$ (dotdashed line) and in $SSM$ (dashed line).   
%The lower line indicate the SM cross section $\sigma_{SM}(M_{Z'},\Delta M)$ 
%for $MQLSM$ mass window~$\Delta M$. 
%Gray filled area denotes the MQLSM cross section uncertainties resulted from the uncertainties 
%in parton distribution functions.
}
\label{fig:gr1}
\end{figure}

The computation of the cross sections (\ref{pp_vs_M}) has been performed with using the set
of parton distribution functions \cite{Alekhin} in the leading order with the fixed flavor number scheme.
The figure~\ref{fig:gr1} shows the cross sections $\sigma_s(M_{Z'},\Delta M)$ of $\mu^+\mu^-$ pair 
production at the LHC energy with account of $Z'$ boson of the $MQLS$ model in comparison with 
$E_6$, $LRM$ and $SSM$ models   
as well as the background cross section $\sigma_{SM}(M_{Z'},\Delta M)$
predicted in the SM (for the mass window of MQLS $Z'$ boson).  
As seen  the signal cross section in all the models (incuding the MQLS model) essentially exceeds 
the background one.  
%in all the mass region reachable at the LHC. 

Comparing the results of the $MQLS$, $E_6$, $LRM$ and $SSM$ models it should be noted that 
the $MQLS$ model predicts the relatively large leptonic 
vector coupling constant $v_{\mu}^{Z_2} \sim -1.1 $ and mainly due to this circumstance 
%the enterring in (\ref{pp_vs_M}) expression $\sum_k I_k(M,S)C^{Z'Z'}_{q_k \mu}$ 
in the $MQLS$ model the value of $I^{Z'Z'}(M,S)$ in (\ref{pp_vs_M})
exceeds the corresponding values 
in $E_6$, $LRM$ and $SSM$ models, which increases the $MQLSM$ cross section. 
On the other hand the width of $Z'$ boson in $MQLS$ model with account of the scalar and fermionic decays 
is predicted to be larger than that in 
$E_6$, $LRM$ and $SSM$ models (for example, at $M_{Z'} = 3 \, TeV$ we obtain 
 $\Gamma_{Z'}/M_{Z'} =5.2\%, \, 0.5\%, \, 2.1\%, \, 3.1\%$ 
in $MQLS$, $E_6$, $LRM$, $SSM$ models respectively), which decreases the $MQLSM$ cross section 
due to decreasing the factor $P_{ij}(M^2)$. 
As a result as we see in the figure~\ref{fig:gr1} the signal cross section in $MQLS$ model still exceeds 
the prediction of $E_6$ model and approximately coincides with those in $LRM$ and $SSM$ models.        

\begin{figure}[htb]
 \vspace*{0.5cm}
%\centerline{\epsfxsize=0.6\textwidth\epsffile{LvLast.eps}}
\centerline{\epsfxsize=0.6\textwidth\epsffile{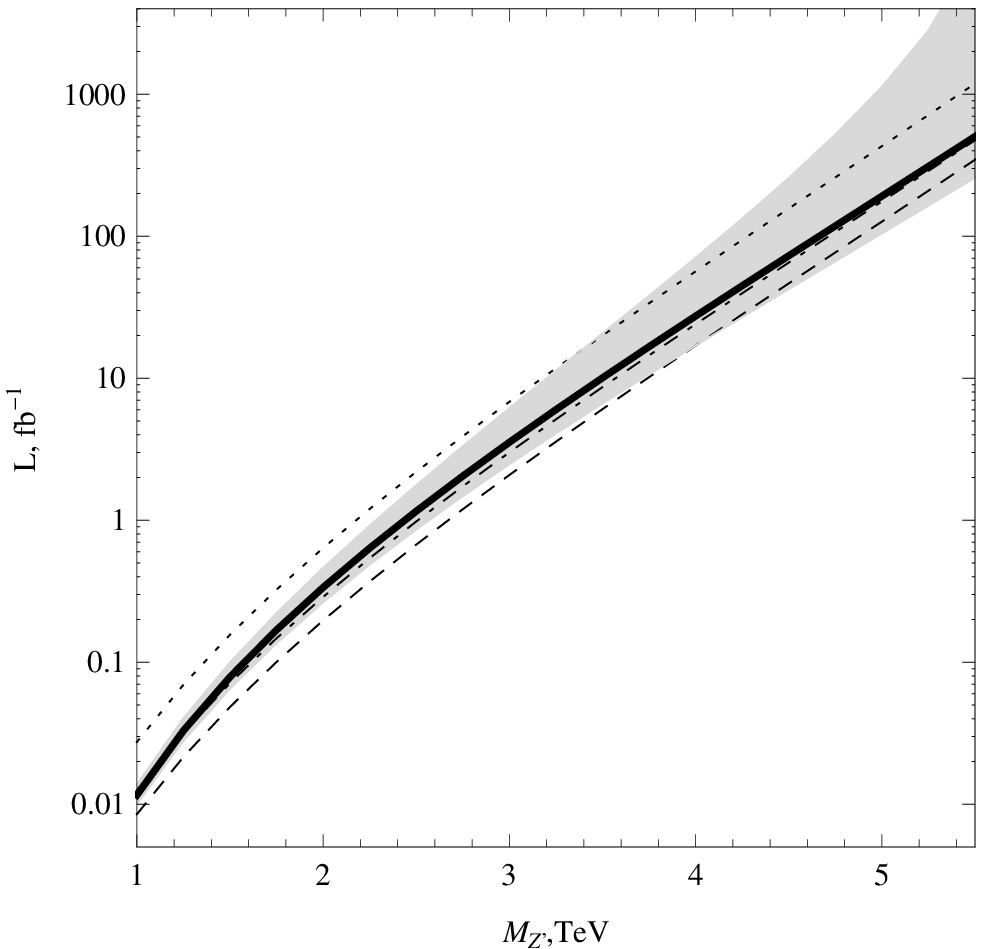}}
%\vspace*{1mm}
\caption{
%Integrated luminosity $L$ needed for $5\sigma$    %($\mathcal{S}'=5$) 
%discovery of $Z'$ boson at the LHC in dependence on $Z'$ boson mass $M_{Z'}$  
%in  $MQLS$ model (solid line), in  $E_6(\psi)$ model (dotted line), 
%in $LRM$ (dotdashed line) and in $SSM$ (dashed line).   
%Gray filled area denotes the MQLSM luminosity uncertainties resulted from the uncertainties 
%in parton distribution functions.
}
\label{fig:gr2}
\end{figure}

The figure~\ref{fig:gr2} shows the integrated luminosity which is necessary for the observation 
of $Z'$ boson of the $MQLS$ model at $5\sigma$ significance ($\mathcal{S}'=5$) 
in dependence on $Z'$ mass in comparison with $E_6$, $LRM$ and $SSM$ models. 
From the figure~\ref{fig:gr2} and using also the formulas~(\ref{pp_vs_M})--(\ref{sigmaSM}) we obtain  
that $Z'$ boson of the $MQLS$ model with masses   
\begin{equation} 
\label{massZ1}
%M_{Z'}<2.47\pm^{0.13}_{0.16} \, TeV, \,\,\, M_{Z'}<3.53\pm ^{0.22}_{0.30} \, TeV, \,\,\, 
%M_{Z'}<4.70\pm ^{0.32}_{0.54} \, TeV  
M_{Z'}<2.44\pm^{0.13}_{0.15}  \, TeV, \,\,\, M_{Z'}<3.50\pm ^{0.22}_{0.29} \, TeV, \,\,\, 
M_{Z'}<4.67\pm ^{0.31}_{0.52} \, TeV  
\end{equation} 
can be observed in $\mu^+\mu^-$ events at the LHC with $5\sigma$ significance at integrated luminosity 
$L= 1 \, fb^{-1}$, $ 10 \, fb^{-1}$, \, $100 \, fb^{-1}$      
with expected numbers of signal (background) events $N_s(N_b) = 3.47 (0.038), \,  3.45 (0.036), 3.34 (0.031)$ 
respectively.   
The unsertainties in~(\ref{massZ1}) are caused by those in parton distribution functions.  

\vspace*{-2mm} 
\begin{table}[htb]
\caption{ 
The upper $Z'$ boson masses in $MQLS$ model observable at the LHC 
in dependence on integrated luminosity $L$ and on significance $\mathcal{S}'\sigma$  
in comparison with  $E_6$, $LRM$, $SSM$ models 
}
\label{tab:MZ1SLM}
\begin{center}
%
%\vspace*{-2mm} 
\hspace*{-15mm}
\begin{tabular}{|c|c|c|c|c|c|c|}
 \hline                                             
               &  \multicolumn{3}{|c|}{$MQLSM $ }     & $E_6$      & LRM        & SSM   \\ \cline{2-7}
$L, \,fb^{-1}$ & $5 \sigma$ & $3 \sigma$ & $1 \sigma$ & $5 \sigma$ & $5 \sigma$ & $5 \sigma$ \\ \hline
  1            & 2.44         & 2.90         & 3.95   & 2.18       & 2.50       &  2.67      \\
 10            & 3.50         & 4.01         & 5.14   & 3.18       & 3.57       &  3.74      \\
100            & 4.67         & 5.19         & 6.38   & 4.28       & 4.72       &  4.89      \\ \hline
\end{tabular}
\vspace*{-3mm} 
\end{center}
\end{table}
The central values in~(\ref{massZ1}) and the corresponding values for $3 \sigma$ and $1 \sigma$ significancies 
in $MQLS$ model are shown in the first columns of the Table~\ref{tab:MZ1SLM}.  
%in comparison with the $5 \sigma$ predictions 
%of $E_6$, $LRM$, $SSM$ models.  
As seen 
%from the table~\ref{tab:MZ1SLM} 
the $MQLSM$ $Z'$ boson with masses 
\begin{equation} 
\label{massZ13s}
M_{Z'} < 5.2 \, TeV
\end{equation} 
can manifest itself in $\mu^+\mu^-$ events at the LHC at integrated luminosity $100 \, fb^{-1}$ 
with $3 \sigma$ significance,  
while for 
\begin{equation} 
\label{massZ11s}
M_{Z'} > 6.4 \, TeV
\end{equation} 
its effect in $\mu^+\mu^-$ events will not exceed $1 \sigma$ i.e. it will be 
practically invisible at the LHC.     

For comparison with the $MQLS$ model the last three columns of the Table~\ref{tab:MZ1SLM} show 
the $5 \sigma$ predictions of $E_6$, $LRM$, $SSM$ models. As seen the central values 
in~(\ref{massZ1}) exceed the corresponding predictions of $E_6$ model and are near 
the predictions of $LRM$ and are slightly below the predictions of $SSM$ model.    
It is worthy to note that the $MQLSM$ $Z'$ boson has some specific features~\cite{PovSmPhAN65} 
which give the possibility in the case of observation of $Z'$ boson to identify its origin.     

In conclusion, we resume the results of the work.

The cross section of the $\mu^+\mu^-$ pair production in $pp$-collisions at the LHC 
is calculated and analysed with acount of the $Z'$ boson induced by the four color 
quark-lepton symmetry ($MQLS$).  
The corresponding $\mu^+\mu^-$ invariant mass spectrum with account of the $Z'$ boson width  
is analysed in dependence on the $Z'$ mass. The mass region for the $MQLS$ $Z'$ boson  
observable at the LHC is found in dependence on the significance and on the integrated luminosity. 
In particular, it is shown that the $MQLS$ $Z'$ boson can be observable at the LHC 
with $3 \sigma$ significance at $L=100 \, fb^{-1}$ up to $M_{Z'} < 5.2 \, TeV$.   
The results are discussed in comparison with predictions of $E_6$, $LRM$, $SSM$ models.

%\vspace{20mm}

%%xxxxxxxxxxxxxxxxxxxxxxx

\newpage

{\Large\bf Figure captions}

\bigskip

\begin{quotation}

\noindent Fig. 1. Cross section $\sigma_{s}(M_{Z'},\Delta M)$ of $\mu^+\mu^-$ pair production at the LHC  
as a function of $Z'$ boson mass $M_{Z'}$  
in  $MQLS$ model (solid line), in  $E_6(\psi)$ model (dotted line), 
in $LRM$ (dotdashed line) and in $SSM$ (dashed line).   
The lower line indicate the SM cross section $\sigma_{SM}(M_{Z'},\Delta M)$ 
for $MQLSM$ mass window~$\Delta M$. 
Gray filled area denotes the MQLSM cross section uncertainties resulted from the uncertainties 
in parton distribution functions.

%\vfill 
\vspace{10mm}
\noindent Fig. 2. Integrated luminosity $L$ needed    
for $5\sigma$    %($\mathcal{S}'=5$) 
discovery of $Z'$ boson at the LHC in dependence on $Z'$ boson mass $M_{Z'}$  
in  $MQLS$ model (solid line), in  $E_6(\psi)$ model (dotted line), 
in $LRM$ (dotdashed line) and in $SSM$ (dashed line).   
Gray filled area denotes the MQLSM luminosity uncertainties resulted from the uncertainties 
in parton distribution functions.

\end{quotation}

\newpage

\vfill 
\centerline{A.~D.~Smirnov, Yu.~S.~Zaitsev, Modern Physics Letters A}

\centerline{Fig. 1}

\newpage

\vfill 
\centerline{A.~D.~Smirnov, Yu.~S.~Zaitsev, Modern Physics Letters A}

\centerline{Fig. 2}

\end{document}